\documentclass[doublecol]{epl2}
\usepackage{graphicx}
\usepackage{subcaption}
\usepackage[american]{babel}
\usepackage{hyperref}	          
\usepackage{epstopdf}
\usepackage{amsmath,amssymb}
\usepackage{mathptmx}

\usepackage[nodisplayskipstretch]{setspace}

\begin{document}

\title{Weyl states and Fermi arcs in parabolic bands}
\author{Mauro M. Doria\inst{1,4}\thanks{\email{mauromdoria@gmail.com}}, Andrea Perali\inst{2,3}\thanks{\email{andrea.perali@unicam.it}}}
\shortauthor{Mauro M. Doria and Andrea Perali}
\institute{
\inst{1} Departamento de F\'{\i}sica dos S\'{o}lidos, Universidade Federal do Rio de Janeiro, 21941-972 Rio de Janeiro, Brazil\\
\inst{2} School of Pharmacy, Physics Unit, Universit\`{a} di Camerino, 62032 - Camerino, Italy\\
\inst{3} INFN, Sezione di Perugia, 06123 - Perugia, Italy \\
\inst{4} Instituto de F\'{\i}sica ``Gleg Wataghin'', Universidade Estadual de Campinas, Unicamp 13083-970, Campinas, S\~ao Paulo, Brasil }
\date{\today}


\abstract{Weyl fermions are shown to exist inside a parabolic band, where the kinetic energy of carriers is given by the non-relativistic Schroedinger equation.
There are Fermi arcs  as a direct consequence of the folding of a ring shaped Fermi surface inside the first Brillouin zone.
Our results stem from the decomposition of the kinetic energy into the sum of the square of the Weyl state, the coupling to the local magnetic field and the Rashba interaction.
The Weyl fermions break the time and reflection symmetries present in the kinetic energy, thus allowing for the onset of a weak three-dimensional magnetic field around the layer. This field brings topological stability to the current carrying states through a Chern number.
In the special limit that the Weyl state becomes gapless this magnetic interaction is shown to be purely attractive,  thus suggesting the onset of a superconducting condensate of zero helicity states.}

\maketitle

\textbf{Introduction}. --  Nowadays many different condensed matter systems display Dirac like properties~\cite{wehling14}, such as the topological insulators~\cite{bernevig06,xia09,hasan10}, graphene~\cite{geim07}, the cuprate superconductors~\cite{damascelli03}, the iron-based superconductors~\cite{liu16} and the Weyl semi-metals~\cite{xu15}. Contrary to normal insulators, topological insulators are conductive at the surface of the material. The presence of a Dirac linear spectrum, directly probed by angle resolved photomission spectroscopy~\cite{wehling14} suggests a profound common ground among such systems. Interestingly many years ago A.A. Abrikosov~\cite{abrikosov98,abrikosov03} addressed the problem of the linear magneto-resistance of the non-stoichiometric silver chalcogenides (Ag$_{2+\delta}$X where (X=Se,Te)~\cite{xu97,husmann02,zhang11}) found at very low magnetic field and explained it through the ad-hoc assumption that fermions obey a  two-component Dirac (Weyl) equation.The Weyl and the Dirac equations have two and four spinorial components, respectively, and in the limit of  zero mass, the Dirac equation decomposes into two independent Weyl ones, as found by Hermann Weyl in the early days of Quantum Mechanics~\cite{weyl29}. From the other side eight years late Conyers Herring~\cite{herring37} discussed  electronic bands in solids that have the same energy and momentum in crystals that lack inversion symmetry. At the point these bands give rise to low-energy excitations with a dispersion relation linear in momentum. These low-energy excitations have been found in many condensed matter~\cite{xu15,lv15,xu17} systems and referred either as massless Dirac fermions or  Weyl fermions.
The presence of the Dirac linear spectrum states is unfolding new concepts in condensed matter physics, such as that of the Weyl semi-metals~\cite{xu15}, whose surface displays Fermi-arcs. The Fermi-arcs have been found before in other compounds, such as in the underdoped high-Tc superconductors~\cite{damascelli03}.
It also leads to the idea of a Lifshitz transition in the Fermi surface~\cite{liu10,chi17}, which is a change of the topology of the Fermi surface without symmetry breaking, e.g., the transformation from a circular to an arc shaped surface. The possibility that surface states can have distinct physics from the bulk has been pointed out long ago by I. Tamm~\cite{tamm32}. Fifty years ago V. Ginzburg~\cite{ginzburg64} argued that in an insulating material the surface layer can acquire metallic character and become superconducting as the pairing (attractive) interaction between carriers may occur only near the surface. From this point of view it is easily acceptable that a topological insulator can become a superconductor at the surface~\cite{zhao15}.
Undoubtly surfaces physics contains new ingredients as compared to the bulk. For instance the vector normal to the surface allows for a new energy invariant in case the carriers have spin, known as the Rashba term~\cite{gorkov01}.\\

In this letter we consider a parabolic band in a single electronic layer. The carriers have spin, charge, and an effective mass in the kinetic energy described by the non-relativistic gauge invariant Schroedinger equation. We show that these  electronic states can display a Dirac linear spectrum by becoming Weyl states~\cite{cariglia14}, which are two-dimensionally confined  in momentum but not in position space.
The Weyl equation is reinterpreted here according to its space and time symmetries and shown to have a cutoff in momentum space, such that  a linear dispersion spectrum exists much above it. In the limit that this cutoff is absent the Weyl fermions are zero helicity states and the magnetic field energy among them is shown to be purely negative, and so, induces attraction among them.
There are two momenta at the Fermi surface, which means that the Fermi surface of the Weyl fermions has the shape of a ring. The presence of a periodic crystallographic potential transforms this ring into Fermi arcs by the onset of a gap at the edge of the first Brillouin zone.
The breaking of the time and the reflection symmetries in the layer  allows for the onset of a local three-dimensional magnetic field with closed streamlines that pierce the layer twice. This field leads to topologically protected states.
The onset of nodal-ring fermions, Weyl fermions, Fermi-arcs  as well as an emergent
Lifshitz transition has been recently discussed for the topological semi-metal TaS~\cite{sun17}.\\

Interestingly the Drude-Sommerfeld model, a cornerstone for the understanding of the physics of metals~\cite{ashcroft76,kittel76},
only considers zero current free fermion states. Then  residual collisions are treated perturbatively in the Boltzmann equation framework.
Therefore the Drude-Sommerfeld model ignores another source of residual interaction among fermions, which is caused by their own magnetic field created by their motion. In the present approach this residual magnetic field interaction  is not neglected and in fact is very important because it brings topological stability to the electronic states through a Chern number. It allows for the onset of inhomogeneous states which become protected from decaying into homogeneous states.
Nevertheless very strong residual collisions can alter this picture and destroy  the closed magnetic field streamlines around the layer.
Therefore the two limits of very long and short mean free paths must be kept in mind as this local magnetic field may survive  in the former but not in the latter case. Thus the starting point here is the hamiltonian for fermions in a layer such that the magnetic energy created by  their kinetic motion is also considered.
\begin{eqnarray}\label{hamilton01}
&& H =  K+F+V,\\
&& K\equiv  \int d^3x \; \frac{1}{2m}|\vec{P}\Psi|^2, \; F \equiv \int d^3x \;\frac{1}{8\pi} \vec h(\Psi)^2, \\
&& V(\vec x + \vec a) = V(\vec x),
\end{eqnarray}
where $\vec P = (\hbar/i)\vec \nabla - (q/c)\vec A$ and  $\vec h =\vec \nabla \times \vec  A$ are the momentum  and the local magnetic field, respectively.
The local magnetic field, $\vec h\left( \Psi \right)$, is obtained by solving Amp\`ere's law,
\begin{eqnarray}
\vec \nabla \times \vec h = \frac{4\pi}{c} \vec J, \quad  \vec J = \frac{q}{2m}\left ( \Psi^{\dag }\vec P \Psi + c.c. \right ).\label{ampere01}
\end{eqnarray}
The potential energy $V$  describes the crystal lattice potential periodic in $\vec a$. \\

The present results stem from a three-term decomposition of the kinetic energy $K$ that generalizes the  Liechnerowitz-Weitzenb\"ock decomposition~\cite{alfredo13}. This decomposition states that for a space endowed with spin, the Laplacian operator can be expressed as the sum of the square of the Dirac (Weyl) operator plus the curvature of the space, which is given by the local magnetic field interaction. However there is a third term, which is the Rashba surface interaction term.
Thus the three-term decomposition of the kinetic energy also opens the way to a direct description of the spin-orbit effects associated to the Rashba term in case of parabolic bands.
The three-term decomposition of the kinetic energy has been applied to the Ginzburg-Landau theory of a layered superconductor suggesting that the pseudo-gap can be a topological state~\cite{alfredo13,alfredo14,cariglia14,doria14,alfredo15}.
This decomposition was also used to explain the  transverse magnetic moment and torque observed by Li et al.~\cite{luli11} in the LaAlO$_3$/SrTiO$_3$ interface~\cite{edinardo16,edinardo17}.\\

\textbf{The Weyl equation and the reflection symmetry}.-- The kinetic energy is invariant under space and time symmetries but  the Weyl equation,
\begin{eqnarray}\label{weyl0}
v_0\vec \sigma \cdot \vec P \Psi  = E_0 \Psi,\quad  \Psi =
\left(
\begin{array}{c} \psi_{\uparrow} \\ \psi_{\downarrow} \end{array} \right),
\end{eqnarray}
where $v_0$ is a velocity parameter, is not. The Weyl equation must be reinterpreted on light that the helicity operator is a pseudoscalar, and so, it cannot be regarded as a hamiltonian  equation, and so $E_0$ cannot be an energy. The helicity operator is the projection of the total angular momentum ($\vec J = \hbar \vec \sigma/2 + \vec x \times \vec P$) along momentum: $\vec J \cdot \vec P = \hbar \vec \sigma \cdot \vec P/2$, where  $\sigma_i$, $i=$1, 2 and 3 are the Pauli matrices that represent spin.
Under the inversion transformation, defined as $\vec x \rightarrow -\vec x$, $\vec \sigma \cdot \vec P \rightarrow -\vec \sigma \cdot \vec P$, as the momentum is a vector, $\vec P \rightarrow -\vec P$, and angular momentum a pseudo-vector, $\vec J \rightarrow \vec J$.
At this point we stress the presence of two views of the Weyl equation, which are being discussed in the literature, referred here as the Abrikosov-Weyl and graphene-Weyl. They were used to solve the quantum linear magneto resistance and to describe the Dirac cone in graphene, respectively. There the Pauli matrices can represent either spin (pseudo vector) or iso-spin (vector), respectively, since for graphene the isospin is associated to hopping between two distinct triangular lattices.
Nevertheless these two interpretations are fundamentally different, as $E_0$ is a scalar in the former and a pseudo-scalar in the latter.
Since most of the materials do not possess intertwined triangular lattices, the graphene-Weyl equation is not generally applicable. From its side the Abrikosov-Weyl equation is generally applicable since it makes no demand about the underlying hopping lattice but $E_0$ is a pseudoscalar whereas the energy must be necessarily a scalar. For this reason we choose the Abrikosov-Weyl equation to study and reinterpret it in the following way.
Consider a layer at $x_3=0$ that sections space into two independent semi-spaces, namely, $x_3<0$ and $x_3>0$.
The inversion symmetry, $\vec x \rightarrow -\vec x$, is equivalent to a rotation of $\pi$ around the $x_3$ axis, $(x_1, x_2, x_3) \rightarrow (-x_1, -x_2, x_3)$ times the reflection symmetry in the layer, $x_3 \rightarrow -x_3$.
Hence inversion and reflection symmetries are basically the same operation since the Weyl equation is rotationally invariant.
The Weyl equation, given by Eq.(\ref{weyl0}), is valid, however it is just able to determine the momentum perpendicular to the layer, $k_3$, as a function of an adjustable  pseudo-scalar parameter, $\alpha_0$, or equally, from the the  scalar parameter $\theta$, defined below,
\begin{eqnarray}\label{weyl0b}
\alpha_0 \equiv \frac{E_0}{v_0}\equiv\frac{x_3}{\vert x_3\vert }  \hbar \theta.
\end{eqnarray}
Hence  in case of no magnetic field ($\vec P = (\hbar/i) \vec \nabla$), the Weyl state is a solution of Eq.(\ref{weyl0}), given by,
\begin{equation}
\Psi(\vec{x}) = \frac{1}{\sqrt{V}} \sum_{\vec k_{\parallel}} c_k e^{i\vec{k_{\parallel}}\cdot\vec{x}} e^{i k_3 x_3 }\begin{pmatrix} 1\\ \displaystyle - i\frac{k_{+}}{k} \frac{x_{3}}{\vert x_{3}\vert } \left (\frac{\theta}{k}-\frac{k_3}{k} \right ) \end{pmatrix}.
\end{equation}
Along the layer we assume periodic boundary conditions, characterized by the unit cell length, $L$ ($A \equiv L^2$).
However the dimensionality of $\Psi$ is associated to a volume, $V=A L_3$, where $L_3$ is an arbitrary length perpendicular to the layer, that later will be related to the velocity of carriers along the layer, $v_0$.
The wave numbers parallel and perpendicular to the layer are $k \equiv \vert \vec k_{\parallel} \vert = \sqrt{k_{+}k_{-}}$, and $k_3= \pm \sqrt{\theta^2-k^2}$, respectively, where $\vec k_{\parallel} = k_1 \hat x_1 + k_2 \hat x_2 $, and $k_{\pm}\equiv k_1\pm i k_2$. The Weyl state annihilation and creation operators with momentum $\vec k_{\parallel}$ are $c_{\vec k_{\parallel}}$ and ${c_{\vec k_{\parallel}}}^{\dag}$, respectively: $\{{c_{\vec k_{\parallel}}}^{\dag}, c_{\vec k_{\parallel}'} \}=\delta_{\vec k_{\parallel},\vec k_{\parallel}'}$ and the Kronecker delta function is defined as $\delta_{\vec k_{\parallel},\vec k_{\parallel}'}=1$ for $\vec k_{\parallel}=\vec k_{\parallel}'$ and  $\delta_{\vec k_{\parallel},\vec k_{\parallel}'}=0$ for $\vec k_{\parallel}\ne \vec k_{\parallel}'$.
In order to comply with the inversion property of the wave number ($\vec k_{\parallel}, k_3$), the $k_3$ free sign is chosen such that,
\begin{equation}\label{k3}
k_3 = i  \frac{x_3}{\vert x_3 \vert} q_3(k), \quad q_3(k) \equiv \sqrt{k^2-\theta^2}>0.
\end{equation}
The states with $k < \theta$ are not considered here because they are travelling waves, thus  not confined to the layer. We only take into account states with $k \ge \theta$ as they evanesce away from the layer and so, are not confined to it.
We reexpress the Weyl solution as,
\begin{equation}\label{psi-a}
\Psi(\vec{x}) = \frac{1}{\sqrt{V}} \sum_{\vec k_{\parallel},\; k \ge \theta} c_k e^{i\vec{k_{\parallel}}\cdot\vec{x}} e^{-q_3 \vert x_3 \vert  }\begin{pmatrix} 1\\ \displaystyle \frac{x_{3}}{\vert x_{3}\vert }\frac{k_{+}}{k} \left (\frac{\theta-iq_3}{k}\right ) \end{pmatrix}.
\end{equation}
A direct inspection of $\Psi$ shows explicit dependency of the product $x_3/\vert x_3 \vert$ times $k_+/k$ which means the breaking of the  time and reflection symmetries independently, respectively, but not their product. Since $\Psi$ evanesces away from  the layer according to Eq.(\ref{psi-a}), fermions  are two-dimensionally confined in momentum but not in position space.
Nevertheless the Fermi surface is in some way spatially confined because
the inert fermions deep inside the surface are three-dimensional ($e^{-k |x_3|}\sim 1.0$ for $k \sim 0$)
while those at the surface are effectively two-dimensional ($\sim e^{-k_F |x_3|} \sim 1.0$ only for $x_3 <1/k_F$).
We notice that absolute two-dimensional confinement is energetically costly according to the Heisenberg principle. Consider, for instance, a layer of atomic thickness, $\delta x_0 \sim 0.1$ nm. There is an intrinsic uncertainty in momentum given by $\delta p \sim \hbar / \delta x_0$, which corresponds to a velocity $\delta v = \delta p /m \sim 10^6$ m/s assuming for $m$ the electron´s mass. Consequently there is an energy $\delta p^2 /2m \sim 3.8$ eV associated to this uncertainty. The present Weyl states are free of this energy cost.\\

Next we address the question about the obtainment of the linear Dirac spectrum from the simplest possible conditions.
We find that the Rashba energy turns into the linear Dirac spectrum for a Weyl state
The Rashba energy is defined as the sum of the Rashba interaction in all points both above and below the layer.
Hence the Rashba energy is proportional to the integral $I$, defined below.
\begin{eqnarray}
I \equiv \int d^2x \; \hat x_3 \cdot\left( \Psi^{\dag}\,\vec \sigma \times \vec P \,\Psi\vert _{x_3=0^-} -  \Psi^{\dag}\,\vec \sigma \times \vec P \,\Psi\vert _{x_3=0^+} \right)
\end{eqnarray}
Under the condition of a Weyl state, the Rashba interaction becomes,
$\Psi^\dag \hat x_3 \cdot \left (\vec \sigma \times \vec P \right)\,\Psi +c.c. = -\hbar \hat x_3 \cdot \vec \nabla\left ( \Psi^\dag\Psi \right)$.
This identity is obtained straightforwardly by firstly expressing the Weyl equation as follows.
\begin{eqnarray}
\vec \sigma \times \vec P \Psi  = i \alpha_0 \vec \sigma \Psi-i \vec P \Psi
\end{eqnarray}
Then $I = \hbar\int d^2x \; \left [\left (\hat x_3 \cdot \vec \nabla\Psi^{\dag}\Psi \right)|_{x_3=0+}- \left (\hat x_3 \cdot \vec \nabla\Psi^{\dag}\Psi\right)|_{x_3=0-}\right ]$.
Next we transform $I$ into a volumetric integration  by means of Gauss' theorem  applied over the volumes above and  below the layer.
This is possible because of the evanescence away from the layer. Then it holds that $I= \hbar  \int d^3x \;  \partial^2 \left ( \Psi^{\dag}\Psi\right)/\partial x_3^2$ and because of the periodicity $L$ along the layer, one finally obtains that,
\begin{eqnarray}\label{iden2}
I= \hbar \int d^3x \; \vec \nabla^2 \left ( \Psi^{\dag}\Psi\right) = \frac{8\hbar}{L_3}\sum_{\vec k_{\parallel}} \; q_3(k) {c_{\vec k_{\parallel}}}^{\dag} c_{\vec k_{\parallel}}.
\end{eqnarray}
We conclude that $\theta$ is a cutoff in the Weyl spectrum for $\theta \ne 0$ and  in the limit $k \gg \theta $ the Dirac cone is obtained. The zero helicity state corresponds to  $\alpha_0=0$ and for it the Dirac linear spectrum is exact for any $k$ since $q_3(k)=k$.

\textbf{The three-term  decomposition of the kinetic energy}. --
The non-relativistic gauge invariant standard kinetic energy admits a distinct but equivalent formulation to that given in Eq.(\ref{hamilton01}), as the sum of three distinct terms, namely, the square of the Dirac operator, the curvature of the space provided by the local magnetic field and the Rashba interaction.
\begin{eqnarray}\label{K02}
&& K = \int d^3x \; \frac{1}{2m}\vert\vec{\sigma}\cdot\vec P\Psi \vert^2-\frac{\hbar}{4m}\vec \nabla \cdot \left[\Psi^\dag\left(\vec \sigma \times \vec P\right)\Psi+c.c. \right] \nonumber\\
&&+\frac{\hbar q }{2 m c}\vec{h} \cdot\left(\Psi^\dag\vec\sigma\Psi\right)
\end{eqnarray}
Then the kinetic energy under the assumption of a Weyl state becomes,
\begin{eqnarray}\label{K03}
K = \int d^3x \; \frac{\hbar^2\theta^2}{2m}\Psi^\dag\Psi +
\frac{\hbar^2}{4m}\nabla^2 \left (\Psi^\dag \Psi \right )+ \frac{\hbar q }{2 m c}\vec{h} \cdot\left(\Psi^\dag\vec\sigma\Psi\right).
\end{eqnarray}
Notice that although the Weyl state breaks the reflection symmetry, the kinetic energy remains unbroken because only the square of the Weyl equation enters on it. An original expression for the current is also obtained, distinct but equivalent to that of Eq.(\ref{ampere01}).
\begin{eqnarray}
\vec J = \frac{q}{2m}\left [ \left(\vec \sigma \cdot \vec D \Psi\right)^{\dag}\vec \sigma \Psi +c.c \right]-\frac{\hbar q}{2m}\vec \nabla \times  \left( \Psi^{\dag}\vec \sigma \Psi\right)
\end{eqnarray}
Imposing a Weyl state through Eq.(\ref{weyl0}), the current becomes,
\begin{eqnarray}
\vec J = \vec J_0-\frac{\hbar q}{2m}\vec \nabla \times  \left( \Psi^{\dag}\vec \sigma \Psi\right), \; \vec J_0 \equiv \frac{q}{m}\alpha_0 \Psi^{\dag}\vec \sigma \Psi
\end{eqnarray}
To determine the magnetic energy we solve Amp\`ere's law and the local magnetic field is found to be,
\begin{eqnarray}\label{hlocal01}
&& \vec h  = -\frac{h q}{m c}\Psi^{\dag}\vec \sigma \Psi + \vec \nabla \times \vec A_0, \\
&& \vec A_0(\vec x) = \frac{4\pi}{c}\alpha_0 \int d^3 x'\; \frac{\Psi^{\dag}(\vec x')\vec \sigma \Psi(\vec x')}{\vert \vec x - \vec x'\vert}
\end{eqnarray}
 For the above local magnetic field it holds that $\vec \nabla \cdot \vec h = 0$ since $\vec \nabla \cdot \left (\Psi^{\dag}\vec \sigma \Psi \right)=0 $ for the Weyl state. Remarkably  the magnetic field streamlines form closed loops that cross the layer twice.
This follows straightforwardly from the expression of $\vec h$ of Eq.(\ref{hlocal01}) using $\Psi$ of Eq.(\ref{psi-a}), which is
well justified in the limit of a very weak field. At this point enters the central assumption that the local magnetic field generated by the motion of the carrier is very small. We define a small parameter $\epsilon$ to quantify this, and since the local magnetic field is quadratic in $\Psi$ it holds that $\Psi \sim O(\epsilon)$ and $h \sim O(\epsilon^2)$. In this way the solution $\Psi$, given by  Eq.(\ref{psi-a}), also applies in presence of a field, but to the lowest order in $\epsilon$.
The magnetic field components parallel to the layer flips sign from one side to the other of the layer, ($\vec h_{\parallel}(O^+)=- \vec h_{\parallel}(0^-)$), whereas the perpendicular one is continuous across it ($h_3(O^+)=h_3(0^-)$).
The conclusion is that the local field streamlines are closed around the layer, and so, bring  stability for the current carrying states since these closed streamlines cannot be spontaneously ruptured unless by the presence of strong scattering.
Nevertheless we emphasize that the topological stability stems from a deeper argument based on the Chern number associated to the mapping of a torus into a sphere. The former is defined by the unit cell within the layer and the latter by the magnetic field unit vector, which is defined in each point of this unit cell. Inhomogeneous states  become stable by belonging to distinct topological classes defined by the integer $Q$, regardless whether their energy is higher than that of the homogenenous state.
\begin{eqnarray}\label{skyrmion}
Q= \frac{1}{4\pi}\int_{x_3=0^+} \big (\frac{\partial \hat
h}{\partial x_1} \times \frac{\partial \hat h}{\partial x_2}
\big)\cdot \hat h \; d^2x.
\end{eqnarray}
To obtain the topological number $Q$ of a given quantum  eigenstate $\vert \Phi >$ of the hamiltonian $H(\Psi)$, one must compute its associated local field, $<\Phi\vert\vec h(\vec x) \vert \Phi>$, whose spatial direction  represented by
$\hat h(\vec x)$. Inhomogeneous solutions have $Q\neq 0$, whereas the homogeneous solution has $Q=0$.\\

Until here we have discussed aspects of the theory $H=K+F$ with a residual but important local magnetic field.
Next we consider the theory $H=K+V$ and consider F residual in order to determine the Fermi surface. Indeed $K$ contains terms of order $O(\epsilon^2)$ and $O(\epsilon^4)$, the latter associated to the local magnetic interaction. In its lowest order  the kinetic energy  in momentum space, is given by,

\begin{eqnarray}
&& K = \frac{\hbar^2}{2mL_3^2} \sum_{\vec k_{\parallel},\; k \ge \theta}E(k){c_{\vec k_{\parallel}}}^\dag c_{\vec k_{\parallel}} + O(\epsilon^4),\label{kin01} \\
&& E(k) \equiv L_3\left [2 \frac{\theta^2}{\sqrt{k^2-\theta^2}}
+4\sqrt{k^2-\theta^2}\right ].\label{kin02}
\end{eqnarray}

\begin{figure}[ht]
\center
\includegraphics[width=\columnwidth]{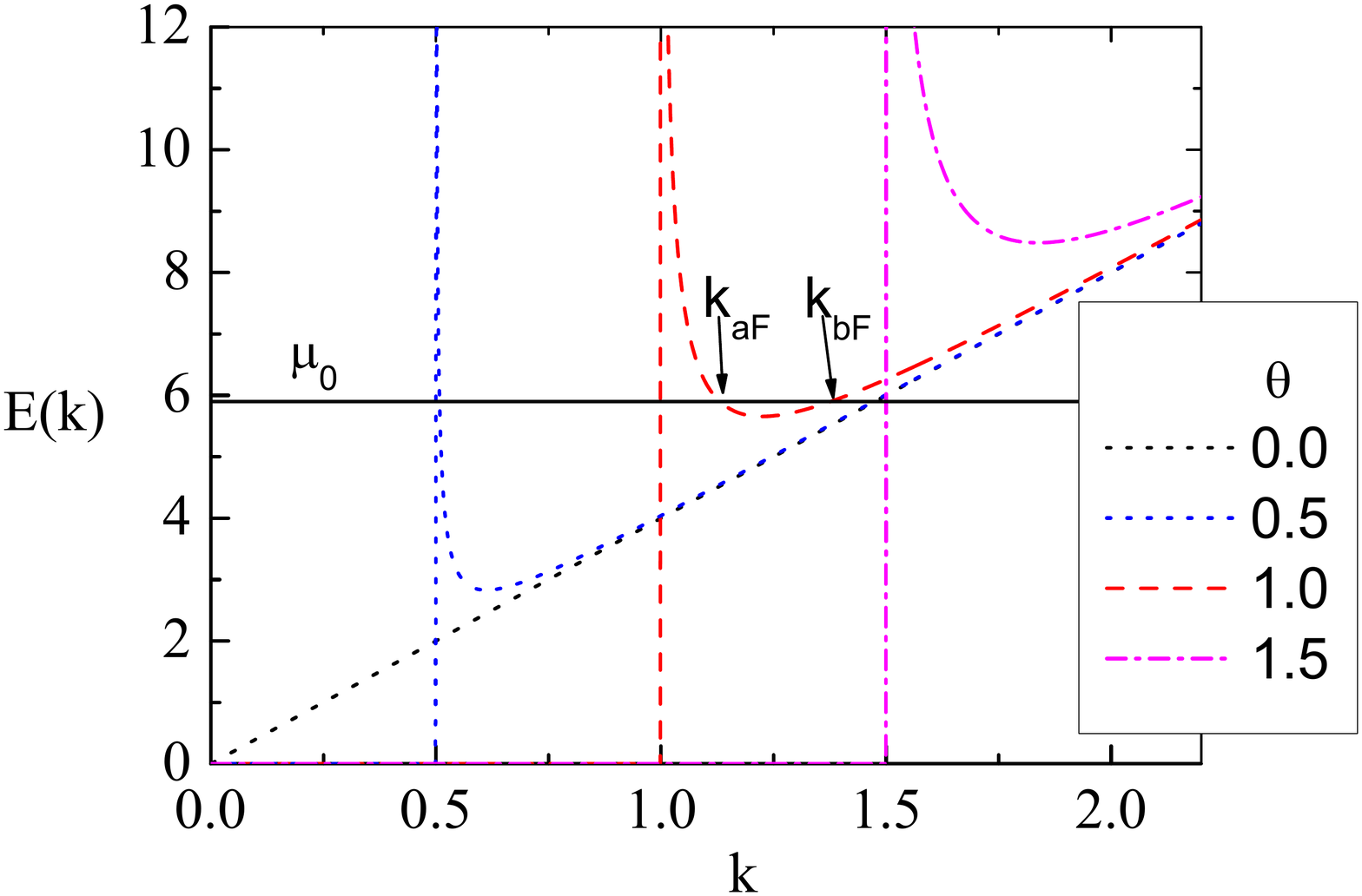}
\caption{The energy described by Eq.(\ref{kin02}) is versus $k$ is shown for a few values of the parameter $\theta$.
The figure shows that the Fermi surface is ring shaped since it possesses  outer and inner radii, called as $k_{aF}$ and $k_{bF}$, as defined in Eq.(\ref{mu0}).
These radius are depicted for the specific case of parameter values, namely, $\theta=1.0$ and $\mu_0=6$. The wavenumbers $k$ and $\theta$ are in  units of $L_3^{-1}$. The dimensionless quantities $\mu_0$ and $E(k)$ must be multiplied by $\hbar^2/2mL_3^3$ to recover the dimension of energy. The length $L_3$ is determined by the in-plane velocity, as given by Eq.(\ref{l3v0}). According to the in-plane velocity suggested in the text, the wave vectors ($k$, $\theta$) and the energies ($E(k)$, $\mu_0$) must be multiplied by $5.0$ nm$^{-1}$ and $24.2$ meV, respectively.}
\label{fig1}
\end{figure}

The limit $k >> \theta$ reveals that the out of layer length, $L_3$ and the in-plane velocity, $v_0$, are related to each other by,
\begin{eqnarray}\label{l3v0}
L_3=2\frac{\hbar}{m v_0}.
\end{eqnarray}
For a typical velocity of $v_0 \sim 10^6$ m/s, such as found in graphene, one obtains that $L_3 \sim 0.2$ nm, assuming for $m$ the electron's mass. Then the energy scale is set by $\hbar^2/2mL_3^3 \sim 24.2$ meV.
The number of particles operator is,
\begin{eqnarray}
N\equiv \int d^3x \; \Psi^\dag \Psi = \frac{1}{L_3} \sum_{\vec k_{\parallel},\; k \ge \theta}\frac{2}{\sqrt{k^2-\theta^2}}{c_{\vec k_{\parallel}}}^\dag c_{\vec k_{\parallel}}.
\end{eqnarray}
The Fermi surface is ring shaped, because the intersection of the  chemical potential with the above dispersion relation, has two solutions $k_{aF}$ and $k_{bF}$,  associated to the outer and inner radii of the ring, which are given by,
\begin{eqnarray}
k_{aF}^2 = \frac{3}{2}\theta^2 + \frac{\mu_0^2}{32} \left [\delta^2 + \delta \right ], \quad k_{bF}^2 = \frac{3}{2}\theta^2 + \frac{\mu_0^2}{32} \left [\delta^2 - \delta \right ], \label{kakb}
\end{eqnarray}
where,
\begin{eqnarray}
&& \delta \equiv \sqrt{1-2\left(\frac{4\theta}{\mu_0} \right)^2},\label{delta}\\
&& \mu_0 \equiv \frac{2mL_3}{\hbar^2}\mu. \label{mu0}
\end{eqnarray}
Therefore $\sum_{\vec k_{\parallel}}= (A/4\pi) \int^{k_{aF}}_{k_{bF}}dk^2$, and one obtains the following expressions for the particle and the kinetic densities.
\begin{eqnarray}
&&\frac{N}{A}=\frac{1}{L_3\pi} \left [ \left(k_{aF}^2-\theta^2 \right)^{1/2}-\left(k_{bF}^2-\theta^2 \right)^{1/2} \right] \\
&& \frac{K}{A}= \frac{\hbar^2\theta^2}{2m}\frac{N}{A}+ \\
&& \frac{\hbar^2}{2m}\frac{2}{3L_3\pi}\left [ \left(k_{aF}^2-\theta^2 \right)^{3/2}-\left(k_{bF}^2-\theta^2 \right)^{3/2} \right]+ O(\epsilon^4)\nonumber
\end{eqnarray}
Fig.~\ref{fig1} shows the function $E(k)$ vs. $k$ for some values of $\theta$ in units of $L_3$. The spectrum is linear and gapless for the zero helicity states ($\theta=0$): $E(k) = 4kL_3$.
The Weyl state is gapped  for $\theta \neq 0$ since the kinetic energy reaches its minimum value for $k_{min}=\sqrt{3/2}\theta$ where $E(k_{min})=4\sqrt{2}\theta$.
A parabolic behavior is found around this minimum, $E(k) = E(k_{0})+L_3( k - k_{0})^2/(2 m^*)+\cdots$, where $m^*=\theta/(24\sqrt{2})$.
Away from this minimum, for $k >> \theta$, the linear spectrum is retrieved, namely, $E(k) \rightarrow 4kL_3$.

The number of particles and kinetic energy densities in the two limits of zero helicity states ($\theta=0$) and of the extremely thin ring shaped Fermi surface ($\delta <<1$), Eqs. (\ref{delta}), (\ref{mu0})), reveal interesting features of the present approach. In case of zero helicity states, $\theta=0$,  one obtains from the above that $N/A=k_F/L_3\pi$ and $K/N=(2/3) (\hbar k_f)^2/2m$ while for the two-dimensional electron gas with a linear dispersion relation $v_0\hbar k$, $N/A=k_F^2/2\pi$ and $K/N=v_0\hbar k_F^3/3\pi$.
Thus these two models are not equivalent, although both rely on the parameters, $k_F$ and $v_0$. Notice that $k_F$ fully determines $N/A$ for the standard  two-dimensional gas, but not for the present theory, while for $K/N$ it is just the opposite. Next we consider an extremely thin ring shaped Fermi surface, obtained in the limit $\delta << 1$. In this limit $\delta$ is proportional to the ratio between the width of the ring and its radius, $\delta\approx (3/8)(k_{aF}-k_{bF})/(k_{0})$. It follows that $N/A= (m\mu/2\pi\hbar^2)\delta$ and $K/N=(1/16)(\hbar \mu_0)^2/(2m)$. Fig.~\ref{fig1} shows the dispersion relation of Eq.(\ref{kin02}) as a function of $k$ for a few values of the parameter $\theta$.
The Fermi surface changes from  a gapped to a gapless regime as the  Weyl  ($\theta \neq 0$) becomes a zero helicity state ($\theta=0$) without any change of symmetry, thus characterizing a Lifshitz transition~\cite{perali12}\\

\begin{figure}[ht]
\center
\includegraphics[width=\columnwidth]{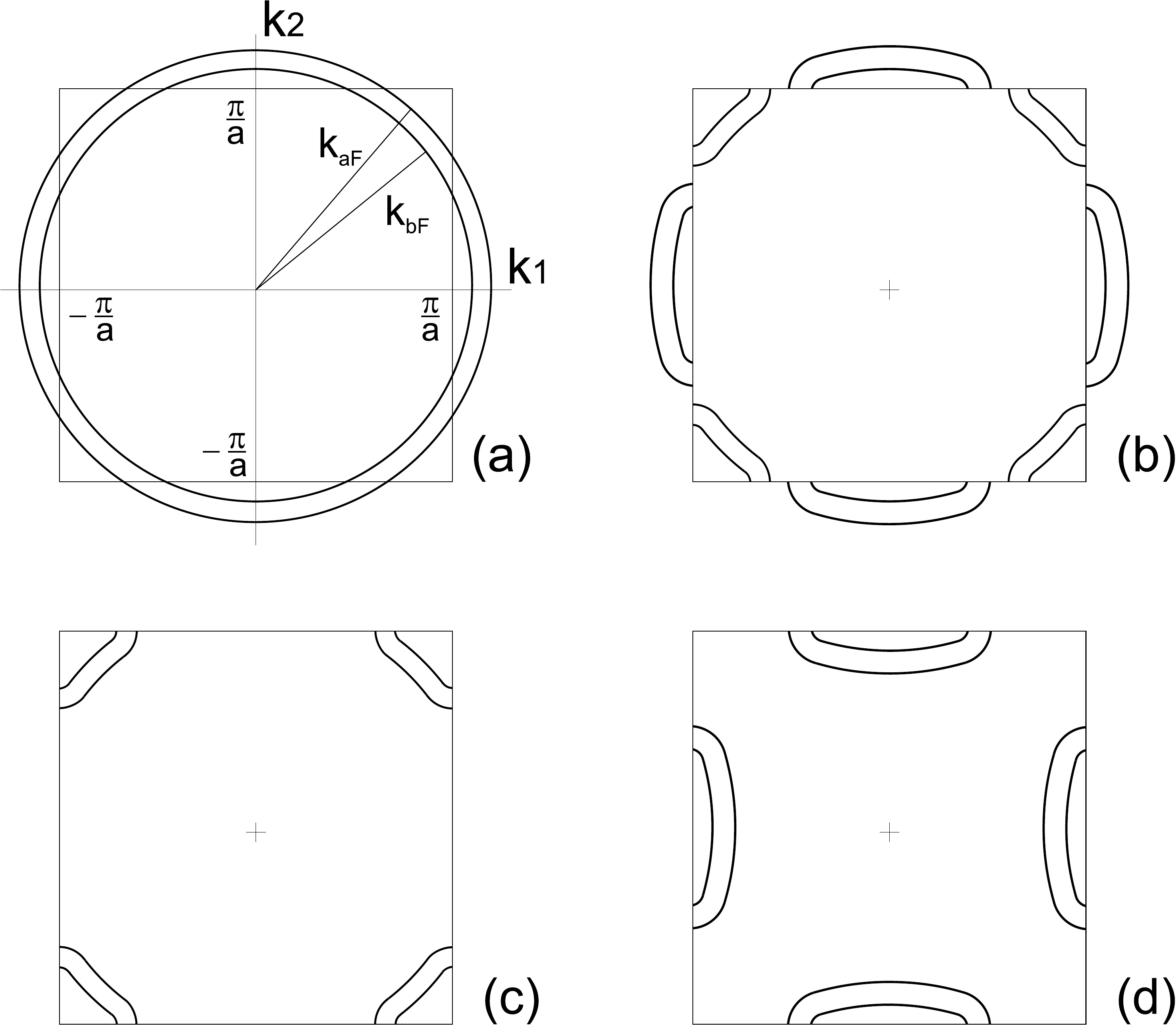}
\caption{The Fermi arcs are a consequence of the crystalline structure into the Weyl dispersion relation, as  shown here.
Figure (a) shows the ring shaped Fermi surface, obtained from Eqs.(\ref{kin01}) and (\ref{kin02}), assumed to fall beyond
the first Brillouin zone of a square lattice with periodicity $a$.
Figure (b) shows the onset of a gap between the lower and upper bands.
Figure (c) shows the Fermi arcs which correspond to the lower band.
Figure (d) shows the upper band reflected inside the first Brillouin zone.}
\label{fig2}
\end{figure}
Next the crystal lattice potential is turned on and we assume that the ring shaped Fermi surface falls beyond the first Brillouin zone brought by a square lattice potential $V$ with periodicity $\vec a$.
Fig.~\ref{fig2}(a) shows both the ring shaped Fermi surface and the first Brillouin zone.
Since we assume that the ring shaped Fermi surface advances into the second Brillouin zone, the result is the onset of two bands, a hole pocket at the Brillouin zone corner and an electron pockets at the Brillouin zone edges, as shown in Fig.~\ref{fig2}(b).
The ring is distorted at the intersection with the edges of the first Brillouin zone by the onset of a gap, according to Bloch's theorem.
The resulting lower and higher energy bands are shown in Figs.~\ref{fig2}(c) and (d), respectively.
In summary the Fermi arcs, shown in Fig.~\ref{fig2}(c),  are an ubiquitous and  robust feature of Weyl states found in  parabolic bands. They are  revealed by the three term decomposition of the kinetic energy.

\textbf{The attractive magnetic interaction of zero helicity states}. --
Zero helicity states have the remarkable property that their magnetic energy is attractive, and so provides a framework for the onset of a condensate. The  zero helicity state is just a  Weyl state with $\theta_0=0$  in Eq.(\ref{weyl0b})).
Interestingly  the local magnetic field of Eq.(\ref{hlocal01}) becomes a local solution, and, the zero helicity states satisfy the following first order equations~\cite{doria10},
\begin{eqnarray}\label{foeq01}
 \vec \sigma \cdot \vec P \Psi  = 0, \quad \mbox{and}, \quad \vec h  = -\frac{h q}{m c}\Psi^{\dag}\vec \sigma \Psi.
\end{eqnarray}
The above first order equations are similar to the  Abrikosov-Bogomolny equations~\cite{abrikosov57,bogomolny76}, which predict vortices, and to the Seiberg-Witten equations~\cite{seiberg94}, which describe four-dimensional massless magnetic monopoles. Each belong to a distinct spin group. The last term of the kinetic energy (Eq.(\ref{K02})) can be expressed as $-\int d^3x\; {\vec h}^2/4\pi$. Added to the field energy gives that,
\begin{eqnarray}\label{hamilton02}
 H=K+F =  \int d^3x \; \frac{\hbar^2}{4m} \vec \nabla^2 \left ( \Psi^{\dag}\Psi\right) -  \frac{1}{8\pi}\vec{h}^2.
\end{eqnarray}
The first term is the Rashba term under the condition of a Weyl state renders the Dirac linear spectrum, as shown in Eq.(\ref{iden2})
Surprisingly the second term is  \revision{negative} and this follows because it  is a sum of the original magnetic energy plus the magnetic energy contained in the kinetic energy. Together they render an  attractive interaction among the fermions that can give rise to  a condensate.
We point out that there is no contradiction to the fact that  $H=K+F$ is positively defined, as seen in Eq.(\ref{hamilton01}).
Hence Eq.(\ref{hamilton02}) must retain this property. Recall that Eq.(\ref{hamilton02}) is an approximate version of  Eq.(\ref{hamilton01}), where the first order equations, Eq(\ref{foeq01}), have been introduced. Thus the negative magnetic interaction should never overcome the linear Dirac spectrum. There is a threshold of positiveness, and to determine it, introduce  $\vec h(\Psi)$ by means of Eq.(\ref{foeq01}) into Eq.(\ref{hamilton02}).
\begin{eqnarray}\label{hamilton03}
H=K+F = \mu_B^2\int d^3x \;\frac{1}{r_e} \vec \nabla^2 \left ( \Psi^{\dag}\Psi\right)-2\pi\left ( \Psi^{\dag}\vec \sigma\Psi\right)^2,
\end{eqnarray}
where we have introduced the Bohr's magneton, $\mu_B=q\hbar/2mc$, which is related to the electron's classical radius by $\hbar^2/4m=\mu_B^2/r_e$, $r_e=q^2/mc\approx 2.8 \, x \, 10^{-6}$ nm.
The lowest wave number is $k \sim 2\pi/L$, thus the condition for a positive spectrum of the above  hamiltonian is simply given by  $L > r_e $, which is always satisfied.\\

\textbf{Conclusion}. -- The Weyl states are discussed in a parabolic band in the context of the three term decomposition of the non-relativistic kinetic energy. The weak magnetic field is the key element to bring  topological stability to the Weyl states. The onset of a condensate of zero helicity states (the pure Dirac cone limit) is envisioned because the magnetic energy becomes purely negative in this limit. The ring shaped Fermi surface of Weyl states transforms into Fermi arcs by the presence of the first Brillouin zone.

\textbf{Acknowledgments}:
The authors thank the colleagues of the Universidade Estadual de Campinas, specially Pascoal Pagliuso and Yakov Kopelevich, for useful discussions. The authors also thank the colleagues involved in the MultiSuper International
Network (http://www.multisuper.org) for exchange of ideas and suggestions for this work. Andrea Perali  acknowledges support by the University of Camerino FAR project CESEMN.

\bibliography{reference}
\end{document}